# Spatial Signal Analysis based on Wave-Spectral Fractal Scaling: A Case of Urban Street Networks

Yanguang Chen, Yuqing Long

(Department of Geography, College of Urban and Environmental Sciences, Peking University, Beijing 100871, P.R. China. E-mail: chenyg@pku.edu.cn)

**Abstract**: For a long time, many methods are developed to make temporal signal analyses based on time series. However, for geographical systems, spatial signal analyses are as important as temporal signal analyses. Nonstationary spatial and temporal processes are associated with nonlinearity, and cannot be effectively analyzed by conventional analytical approaches. Fractal theory provides a powerful tool for exploring complexity and is helpful for spatio-temporal signal analysis. This paper is devoted to researching spatial signals of geographical systems by means of wave-spectrum scaling. The traffic networks of 10 Chinese cities are taken as cases for positive studies. Fast Fourier transform and least squares regression analysis are employed to calculate spectral exponents. The results show that the wave-spectral density distribution of all these urban traffic networks follows scaling law, and the spectral scaling exponents can be converted to fractal dimension values. Using the fractal parameters, we can make spatial analyses for the geographical signals. The analytical process can be generalized to temporal signal analyses. The wave-spectrum scaling methods can be applied to both self-similar fractal signals and self-affine fractal signals in the geographical world.

## 1. Introduction

Signal processing and analysis is important for exploring complex systems, especially when the systems are treated as black boxes. Many mathematical methods can be employed to make signal analyses. A signal is a carrier of information, which can be described by Shannon entropy (Shannon, 1948). Entropy can be treated as one of complexity measures (Cramer, 1993; Pincus, 1991). Hausdorff dimension proved to be equivalent to information entropy (Ryabko, 1986). A number of mathematical and empirical relationships have been found between entropy and fractal dimension (Chen *et al*, 2017; Sparavigna, 2016; Zmeskal *et al,* 2013). Based on box-counting method, normalized fractal dimension was proved to equal normalized information entropy (Chen, 2020). Thus, fractal dimension can be employed to reveal hidden information in signals. Among various fractal-based signal processing and analysis, the most significant ones include reconstructing phase space, rescaled range analysis (R/S analysis), wavelet analysis, and spectral analysis. Spectral analytical processes include power spectral analysis for time series and wave spectral analysis for spatial series. Spectral exponents has been demonstrated to connect fractal dimension, and are associated with the Hurst exponent in R/S analysis (Feder, 1988; Liu and Liu, 1993; Takayasu, 1990).

Generally speaking, signals are represented by time series coming from physical systems or human systems. In fact, signals fall into three categories. The first is temporal signals, which can be reflected by time series data. The second is spatial signals, which can be reflected by spatial data. The third is hierarchical signals, which can be reflected by cross-sectional data indicative of rank-size distribution. Different signals represent different aspects of natural and social systems, and can be characterized by different parameters. The hierarchical signals can be regarded as generalized spatial signals since cross-sectional data are based on spatial random sampling. Many mathematical methods can be employed to analyze signals based on time series, including moving average (MA) analysis, auto-regression (AR) analysis, auto-regressive moving average (ARMA) analysis, power spectrum analysis, and R/S analysis. Under certain conditions, the methods for time series can be properly improved and applied to spatial signal analysis (Brockwell and Davis, 1998). For example, MA and AR can be replaced by spatial autocorrelation analysis and spatial auto-regression analysis, power spectrum analysis can be replaced by wave-spectrum analysis. This paper is devoted to modelling and analyzing spatial signals for geographical systems such as cities. The spatial analysis can be combined with time series analysis. The rest parts are organized as follows. Firstly, the models are presented; secondly, case studies are made by taking 10 Chinese cities as examples; finally, the main points are summarized to conclude this work.

## 2. Models and signals

### 2.1 Spatial signals and urban density models

The similarities and differences between time signals and spatial signals should be clarified first of all. The signals based on time series may be stationary, or may be nonstationary. In contrast, the data series for spatial signals are usually nonstationary. In many case, spatial signals take on certain trends. If the trends can be modeled by a function with characteristic lengths, it belongs to simple signals, and can be addressed by conventional mathematical methods. For example, the spatial series of urban population density follows Clark's law, a negative exponential model. This model bears a scale parameter which represents characteristic length. It is simple and does not need fractal geometry and scaling analysis. On the contrary, if the trends cannot be modeled by a function with characteristic scales, or the trends follows power law, it belongs to complex signals and cannot be dealt with conventional mathematical methods. In this case, fractal-based mathematical methods may be useful for researching this type of signals. For example, traffic network density of a city based on digital maps or remote sensing images can be modeled by Smeed's law (Smeed, 1963). This is an inverse power law, which bears no parameter representing characteristic length. It is complex and needs fractal parameters to describe.

Inverse power function represents the universal trends appearing in spatial signals. Based on this type inverse power law, we can make wave-spectrum analysis. The wave spectrum exponent is associated with fractal dimension in theory (Chen, 2008; Liu and Liu, 1993). The standard form of Smeed's model can be expressed an inverse power law (Batty and Longley, 1994; Chen, 2013)

$$\rho(r) = \rho_1 r^{-\alpha} = \rho_1 r^{D-d}, \tag{1}$$

where $\rho(r)$ denotes the traffic network density at the distance $r$ from the center of city ($r$=0), $d$=2 is the Euclidean dimension of embedding space, $D$ is the fractal dimension, $\rho_1$ refers to traffic network density near city center, and $\alpha=d-D$ to the scaling exponent of the urban traffic network (Chen *et al*, 2019). This fractal dimension represents a type of local dimension, and is termed radial dimension li literature (Frankhauser, 1998; Frankhauser and Sadler, 1991; White and Engelen, 1993). Smeed's model can be treated as a basis for spatial signal analysis. The central density for $r$=0 can be defined as $\rho(0)=\rho_0$. If the spatial distribution of urban density is well described with equation (1), we can calculate the fractal dimension directly using Smeed's model. Thus the wave-spectrum scaling analysis is unnecessary. However, if a city takes on self-affine growth rather than self-similar growth, or the spatial patterns of traffic networks are incomplete due to administration boundaries, we have to make use of wave spectral analysis to address geographical signals.

## 2.2 Fourier transform and wave-spectral scaling analysis

Fourier transform and wave-spectrum scaling relation can be employed to make fractal-based spatial signal analyses for urban evolution. By using the similarity principle of Fourier transform, we can derive the wave-spectral scaling relation between wave number and wave spectral density (Chen, 2010). Using the wave-spectrum scaling relation, we can estimate the fractal dimension of the spatial signals of urban patterns (Chen, 2013; Chen, 2019). Based on equation (1), a density-density correlation function of urban form, $C(r)$, can be expressed as

$$C(r)=\int_{-\infty}^{\infty}\rho(r)\rho(x+r)\mathrm{d}x\,, \tag{2}$$

where $\rho(x)$ refers to the urban density at the distance $x$ from the center of city ($x$=0), and $\rho(x+r)$ to the density at the distance $r$ from the first point at $x$. Equation (2) is a point-point spatial autocorrelation function (Chen, 2013). Suppose that $x$ equals 0, that is, one point is fixed to the city center. Based on the inverse power-law distribution, the point-point correlation function will be reduced to a one-point correlation such as

$$C(r)=\rho(0)\rho_1 r^{D_f-d}\propto r^{-(d-D_f)}, \tag{3}$$

where $D_f$ denotes fractal dimension, which is a type of radial dimension, and $d$=2 is a Euclidean dimension. Applying Fourier transform to equation (3) yields (Takayasu, 1990)

$$F(k)=\int_{-\infty}^{\infty}C(r)e^{-2\pi kir}\mathrm{d}r=2\rho_0\rho_1\int_{0}^{\infty}r^{D_f-d}e^{-2\pi kir}\mathrm{d}r\,, \tag{4}$$

where $F(k)$ denotes the image function of the spatial correlation function, i.e. $C(r)$, and $k$ is wave number. Equation (3) can be simplified as $\mathscr{F}[C(r)]=F(k)$, where $\mathscr{F}$ denotes Fourier operator. Then, applying scaling transform to equation (4) yields

$$F(\zeta k)=\frac{1}{\zeta^{D_f-d+1}}2\rho_0\rho_1\int_{0}^{\infty}(\zeta r)^{D_f-d}e^{-2\pi ki(\zeta r)}\mathrm{d}(\zeta r)=k^{d-D_f-1}F(\zeta k)\,, \tag{5}$$

where $\zeta$ refers to scale factor of scaling transform. The solution to the functional relation, equation (5), is a power law as follows

$$F(k)=F_0k^{-(1+D_f-d)}=F_0k^{-\alpha}, \tag{6}$$

where spectral exponent $\alpha=1+D_f-d$. It is in fact Fourier scaling exponent. If the spatial analysis is made in 2-dimension space, then Euclidean dimension $d$=2, and thus we have

$$\alpha=1+D_f-d=D_f-1. \tag{7}$$

Correspondingly, fractal dimension is

$$D_f=\alpha+1. \tag{8}$$

If we can obtain the spectral exponent by using the Fourier scaling relation, equation (6), then we will get the estimated value of fractal dimension of spatial signals.

Fractal dimension is a spatial characteristic quantity and different types of fractal parameters indicate different spatial properties. In geographical world, there are corresponding relationships between temporal processes, hierarchical structure, and spatial patterns (Batty and Longley, 1994; Harvey, 1969). Time series analysis can be turned into spatial series analysis and *vice versa* (Chen, 2011). In fact, for geographical systems, different types of signals correspond to different types of geographical space (Table 1). Time series reflect phase space, spatial series represents real space, and hierarchal series reflect order space (Chen, 2014). Different spaces have different fractal dimensions. Real space can be described with box dimension and radial dimension, phase space can be described with correlation dimension, and order space can be described with similarity dimension. The fractal-based spatial signal analysis is associated with radial dimension, which can be measured and calculated with cluster growing method, or radius-number scaling method (White and Engelen, 1993). The spatial signals are extracted by means of concentric circles. For urban form and transportation networks, concentric circles can be drawn with the center of a city as the center (Figure 1).

**Table 1 Three types of signals and uses for scientific studies**

| Type | Data | Object | Parameter | Complexity | Geographical space |
|---|---|---|---|---|---|
| **Temporal signal** | Time series data | Dynamic evolution | Hurst exponent | Time lag | Phase space |
| **Spatial signal** | Spatial data | Spatial distribution | Fractal dimension | Spatial dimension | Real space |
| **Hierarchical** | Cross- | Rank-size | Zipf | Interaction | Order space |
| **signal** | sectional data | distribution | exponent | | |

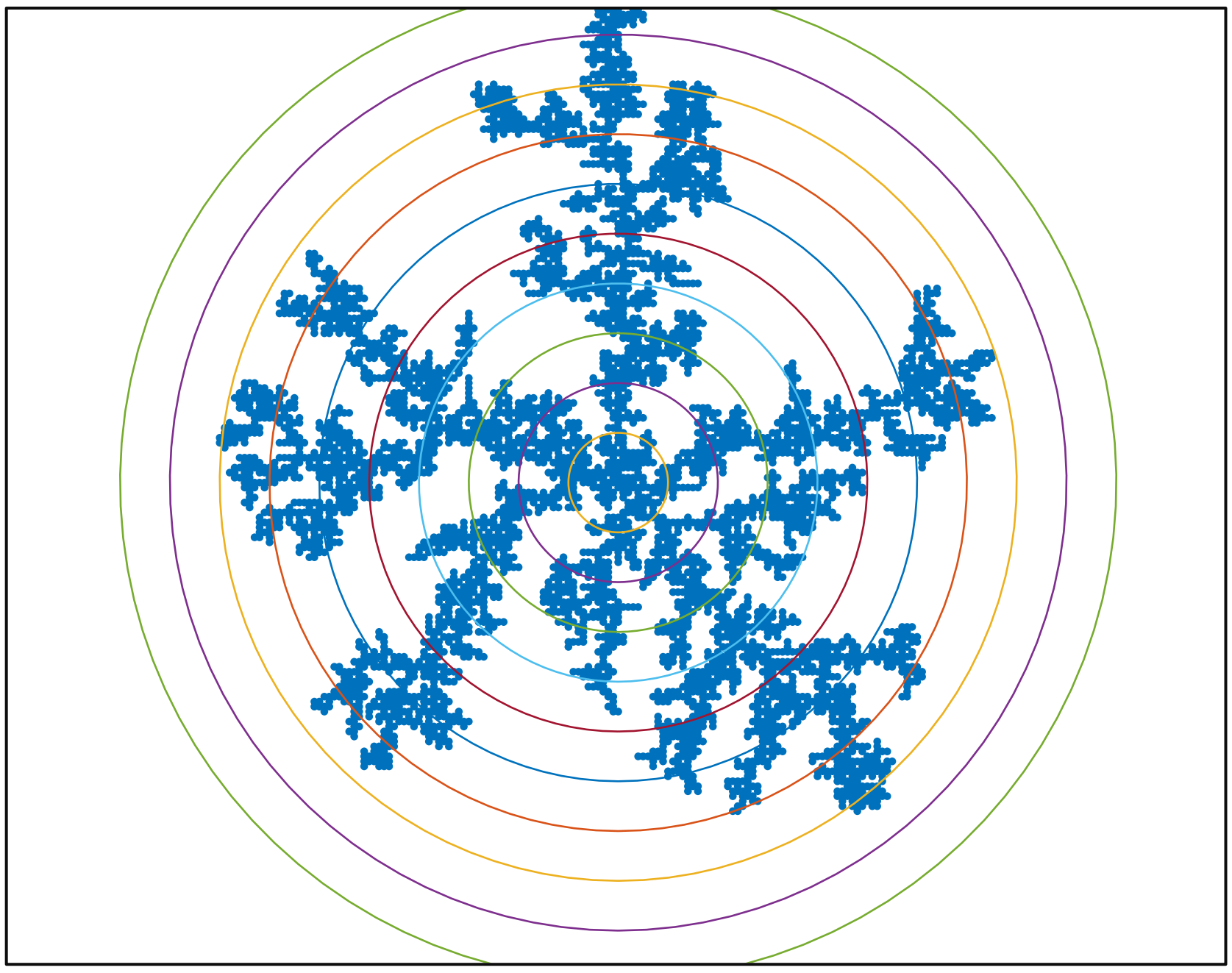

**Figure 1 A sketch map for radial scaling method based growing cluster of cities**

**Note**: The cluster represents diffusion-limited aggregation (DLA) model, which was often employed to simulate urban growth in literature (Batty and Longley, 1994). This cluster pattern was generated by Matlab program.

# 3. Empirical analysis

## 3.1 Data and methods

The methods illustrated above can be applied to 10 cities of China to make case studies. The urban road network dataset is derived from Chinese digital navigation map in 2016 (http://geodata.pku.edu.cn), including freeways, arterials, and collectors. Traffic networks of Chinese cities bear fractal nature, and the fractal dimension can be estimated by means of box-counting method (Long and Chen, 2019a). Box dimension is a kind of global parameter (Frankhauser, 1998). Now, we want to know the local fractal parameter, i.e., radial dimension. Taking the central business district (CBD) of each city as the center, we create 0.5-kilometer to 50-kilometer buffer rings in an increasing, step-wise manner. The ring areas outside the city boundary are excluded. Then, road lengths in each concentric ring are measured by ArcGIS 10.2 (Figure 2). As indicated above, to extract spatial signals of urban traffic networks, we can draw concentric circles with urban center as the center. The circle is numbered as $j$=0, 1, 2, …, $N$-1, where $j$=0 represents the center of a city, and $N$ refers to the total number of circles. In our examples, the number of circles is 100. The spatial signal of urban street lines can be expressed as

$$x(r) = \Delta L(r) = L(r_{j+1}) - L(r_j), \tag{9}$$

where $x(r)$ denotes spatial signals, Δ is difference operator, $L(r)$ is the cumulative length of street lines around city center, $L(r_j)$ is the cumulative length of street lines within the $j$th circle. Therefore, $\Delta L(r)$ refers to the total length of street lines between the $j$th circle and the ($j$+1)th circle. The zone between any two adjacent circles can be regarded as a ring. The average density of each ring can be defined as

$$\rho(r) = \frac{\Delta L(r)}{\Delta A(r)} = \frac{L(r_{j+1}) - L(r_j)}{\pi r_{j+1}^2 - \pi r_j^2}, \tag{10}$$

where $A(r)=\pi r^2$ denotes the area within the circle with radius $r$.

The parameter estimation of any model needs certain algorithms. The wave-spectral analysis need two algorithms: one is fast Fourier transform (FFT), and the other is ordinary least squares (OLS) method. The former is used to obtain wave spectrum, and the latter is used to estimate spectral exponent. Application of a mathematical methods to concrete problems involves what is called three worlds: *real world*, *mathematical world*, and *computational world* (Casti, 1996). In fact, the computational world can be treated as the bridge between the mathematical world and the real world. Equation (4) describes Fourier transform in pure theoretical form, that is, it is defined in the mathematical world. For the observational data from the real world, Fourier transform should be replaced by an algorithm termed FFT. Data processing and algorithm application is implemented in the computational world (Chen, 2019). Applying FFT to the urban density series, we have

$$\rho(r) \xrightarrow{\text{FFT}} F(k) = a + bi, \tag{11}$$

where $F(k)$ is the result of FFT, expressed by complex numbers, $a$ is the real part of the complex number, $b$ is the imaginary part, and $i=(-1)^{1/2}$ is the sign of the complex number. Thus the wave spectrum can be defined as

$$F^*(k) = \left(\frac{|F(k)|^2}{N}\right)^{1/2} = \left(\frac{F(k)\overline{F(k)}}{N}\right)^{1/2} = \left(\frac{a^2+b^2}{N}\right)^{1/2}, \tag{12}$$

Based on equation (12), equation (6) can be substituted with

$$F^*(k) = F_0^* k^{-\alpha}, \tag{13}$$

where the asterisk "*" denotes an equivalent redefinition of variable $F(k)$ and parameter $F_0$ for simplification. The redefinition does not influence the computational results and analytical process of spatial signals. Taking natural logarithm of both sides of equation (13) yields a linear equation as below

$$\ln F^*(k) = \ln F_0^* - \alpha \ln k\,. \tag{14}$$

Thus, by using the OLS method, we can estimate the values of the spectral exponent. In the light of equation (8), the spectral exponent values can be converted into fractal dimension values easily.

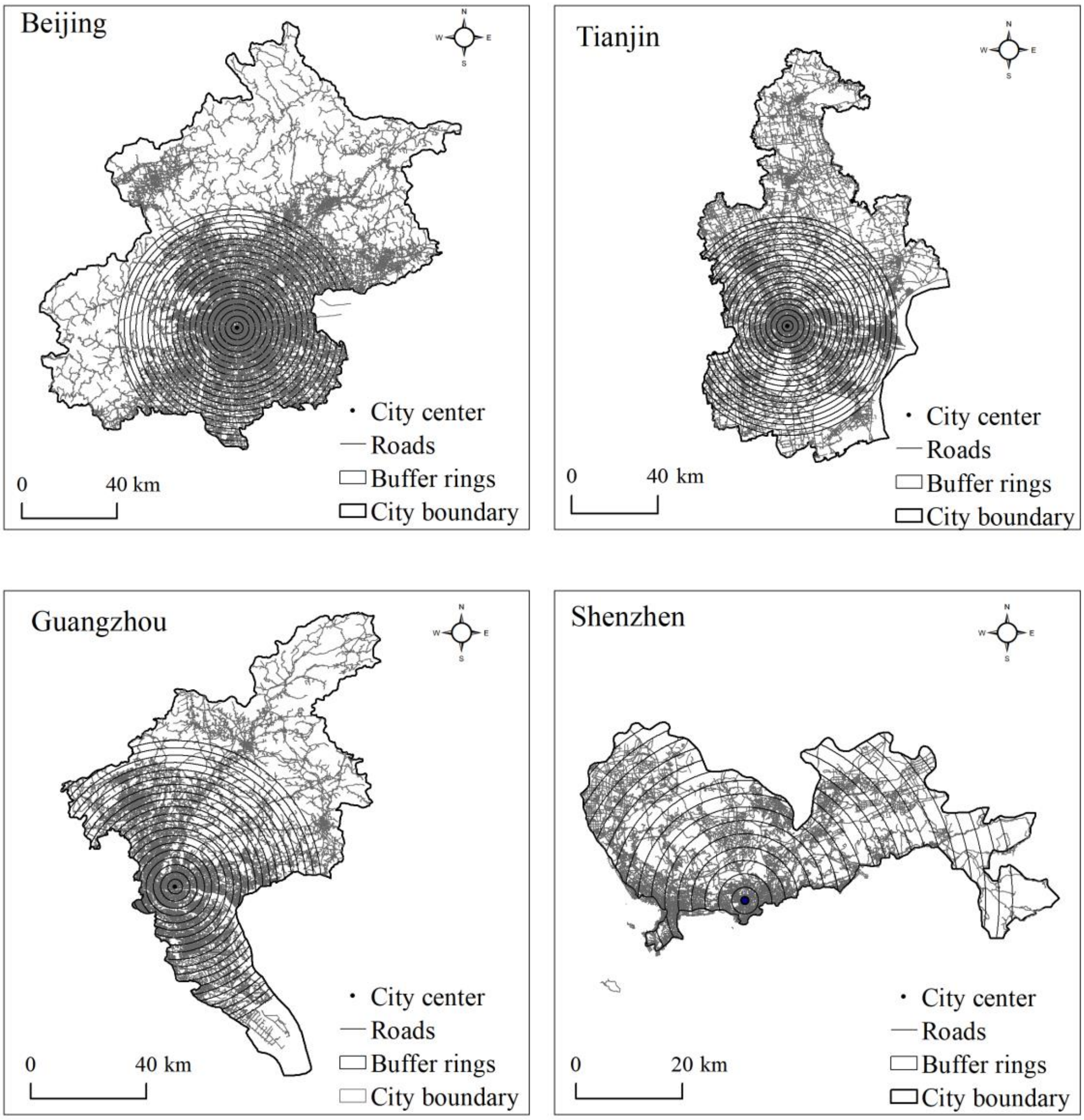

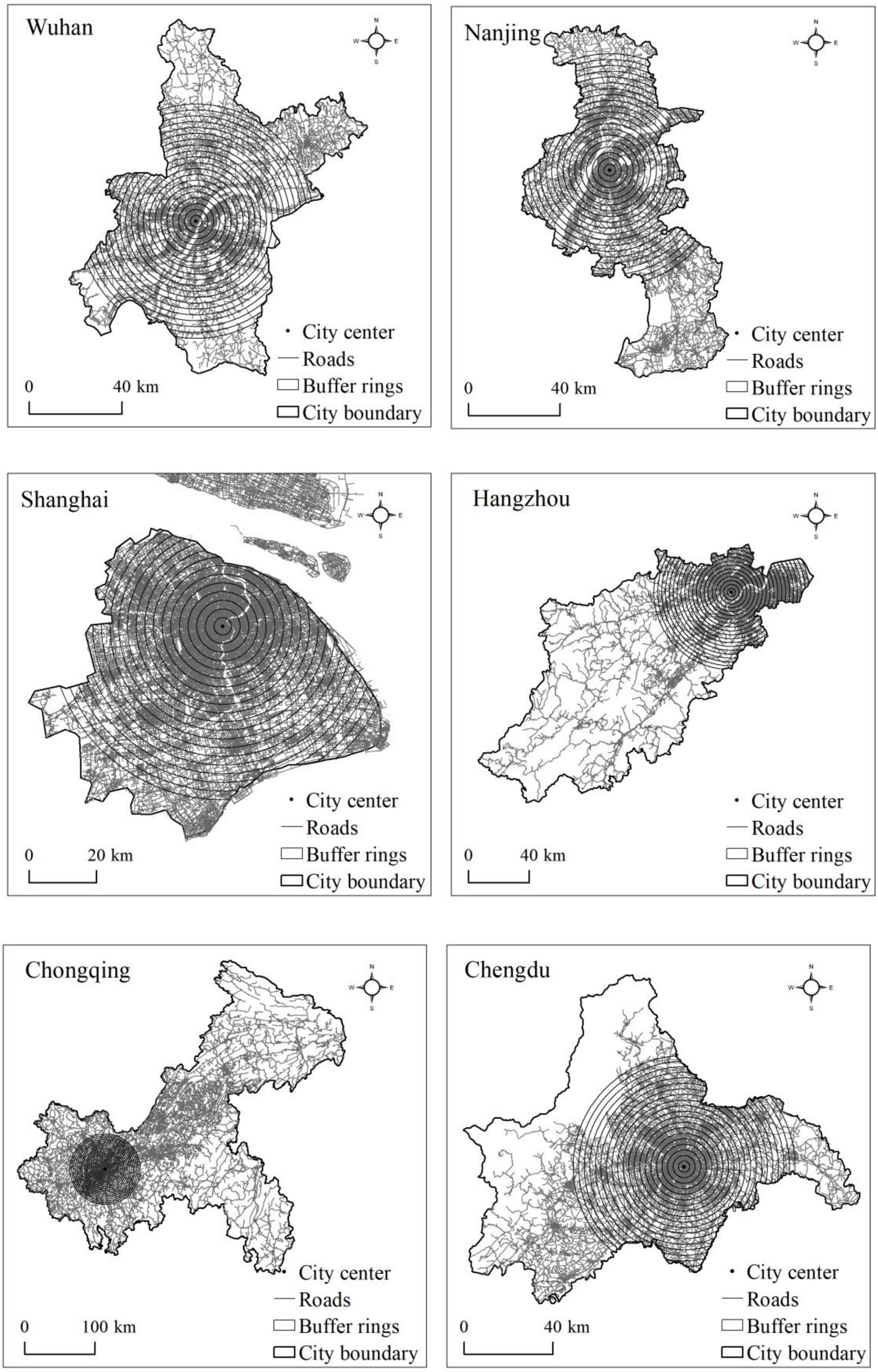

**Figure 2 Traffic networks of 10 Chinese cities based on administrative boundaries in 2016**

**Note**: The data extraction for spatial signals is confined by each municipal boundaries, so that the spatial information is not complete. In this case, it is improper to estimate the fractal parameter using simple fractal model. It is advisable to apply wave-spectrum scaling analysis to these spatial patterns.

## 3.2 Results and analysis

The extraction method of spatial signal depends on the research objective. Suppose that we want to investigate urban growth by using traffic networks comprising streets and roads. Thus, the system of concentric circles can be employed to measure and calculate the total lengths of routes within certain zones. The formula is equation (9). The length series of streets and roads form the origin spatial signals of urban traffic networks. Where trend is concerned, these signals take on unimodal curves (Figure 3 shows two examples). However, we cannot get too more geographical information from these curves. Then, utilizing equation (10), we can turn the network length into network density between two adjacent circles. If the traffic network density follows standard power law, then we can estimate fractal dimension directly. However, due to incomplete patterns of urban network based on administrative boundaries as well as self-affine growth, the network densities come between exponential distribution and logarithmic distribution (Figure 4 shows two examples). In this case, Smeed's model cannot be used to estimate fractal dimension effectively. It is necessary to make wave-spectral analysis using equations (11), (12), (13), and (14).

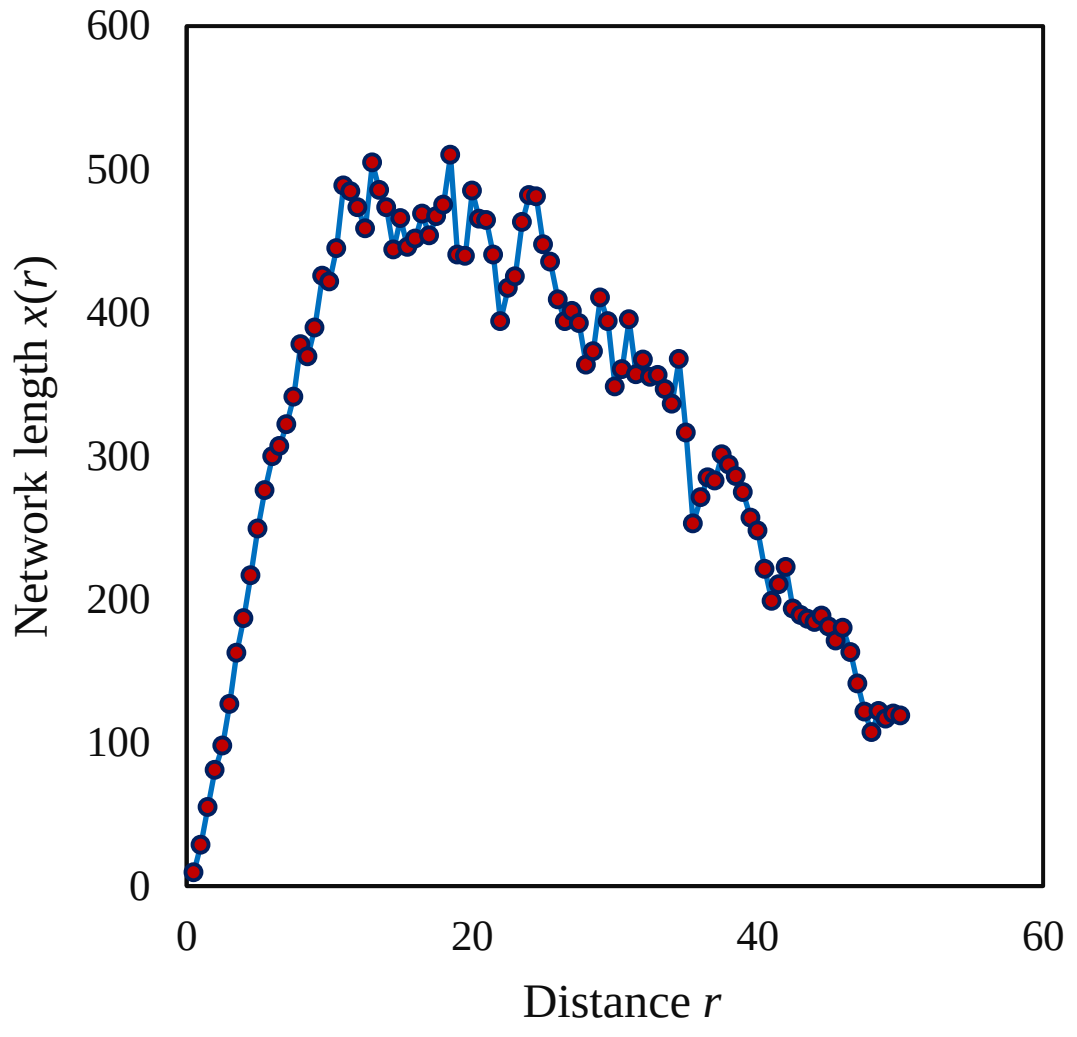

a. Beijing

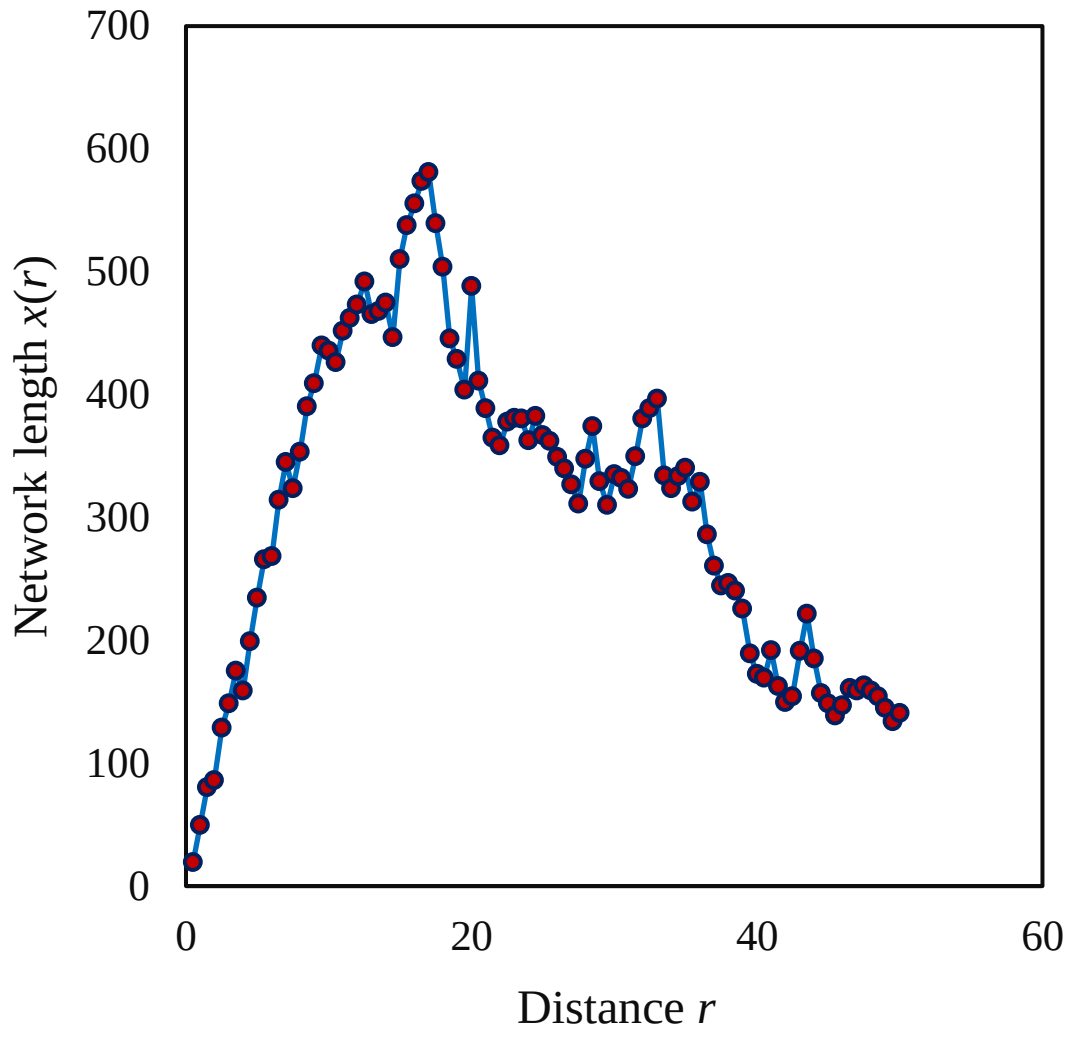

b. Shanghai

**Figure 3 The spatial signals for traffic network of Beijing and Shanghai cities (2016)**

**Note**: These spatial signals represent urban traffic lines extracted through concentric circles and equation (9). Each value reflect a total length of streets and roads falling in two adjacent circles.

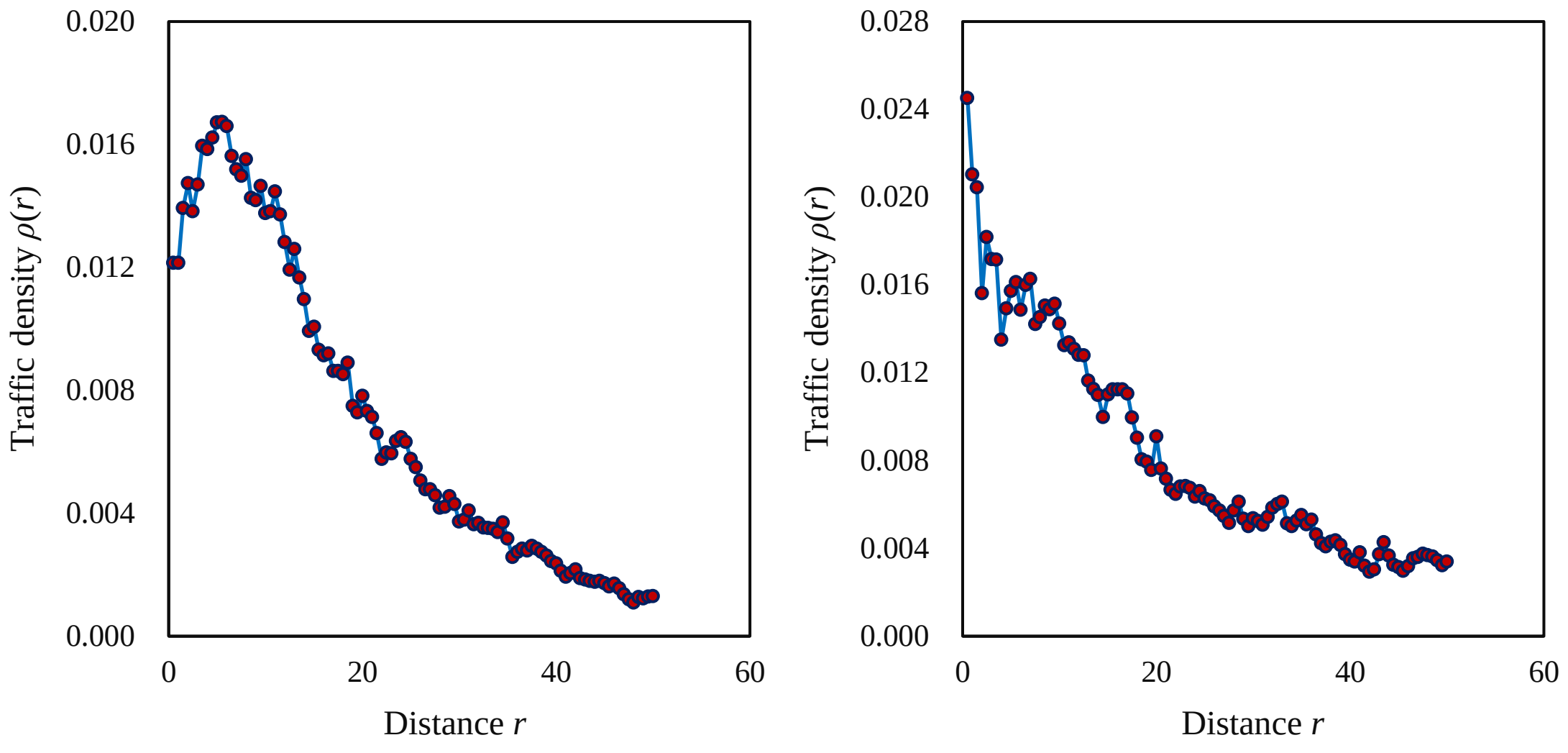

a. Beijing b. Shanghai

**Figure 4 The patterns of traffic density distributions of t of Beijing and Shanghai cities (2016)**

**Note**: The spatial signals can be turned into density curves of urban traffic networks by using equation (10). If the density data could be fitted to an inverse power law, we should estimate fractal dimension directly by means of equation (1). Due to the curves depart from power function, we had to employ the wave-spectrum scaling analysis to estimate fractal dimension.

Based on spatial series of traffic network density, wave-spectrum scaling relations can be examined easily. Using mathematical software Matlab or even MS Excel, we can make Fourier transform for each density series. The wave number can be defined as

$$k = \frac{j}{N}, \tag{15}$$

where $j$=0, 1, 2, …, $N$-1. According to FFT, the number $N$ must satisfy the following condition

$$N = 2^{j+1}. \tag{16}$$

Otherwise, we have to remove a number of numbers at the beginning or at the end of a sequence, or add a set of 0 at the end, so that the length of the sequence becomes the integer power of 2. The circle number of each city is 100. We either add 28 zeros at the end of the sequence so that the length of sample path become 128 ($N$=$2^7$), or remove the last 36 data points so that the sample path length become 64 ($N$=$2^6$). The former way (adding 0) is better than the latter way (removing partial data). Compared with deleting 36 data points, adding 0 at the end of the sequence results in less geospatial information loss. After FFT of density sequence, we can compute wave spectrum by means of equation (12). Then, by using the least square regression based on equations (13) and (14), we can make models for wave spectrum scaling relation. For example, for Beijing's traffic network, the model is

$$\hat{F}^*(k) = 0.0002092k^{-0.9635}. \quad (17)$$

The goodness of fit is about $R^2$=0.8543, the spectral exponent is $\alpha$=0.9635. Thus the fractal dimension is estimated as $D_f$=1.9635. For Shanghai's network density, the model is

$$\hat{F}^*(k) = 0.0005876k^{-0.7707}. \quad (18)$$

The goodness of fit is about $R^2$=0.8570, the spectral exponent is $\alpha$=0.7707. Thus the fractal dimension is estimated as $D_f$=1.7707. The traffic networks of other cities can be dealt with by analogy (Table 2, Figure 5).

**Table 2 Wave spectrum exponents of urban density and corresponding fractal dimension values (2016)**

| Position | City | Coefficient $F_0^*$ | Spectral exponent $\alpha$ | Goodness of fit $R^2$ | Fractal dimension $D_f$ | Fractal dimension $D_s$ |
|---|---|---|---|---|---|---|
| **North** | Beijing | 0.0002092 | 0.9635 | 0.8543 | 1.9635 | 1.5365 |
| | Tianjin | 0.0005071 | 0.7567 | 0.9236 | 1.7567 | 1.7433 |
| **South** | Guangzhou | 0.0003429 | 0.9107 | 0.9152 | 1.9107 | 1.5893 |
| | Shenzhen | 0.0003437 | 0.9838 | 0.8525 | 1.9838 | 1.5162 |
| **Central** | Wuhan | 0.0002729 | 0.8884 | 0.9402 | 1.8884 | 1.6116 |
| **South-east** | Nanjing | 0.0005069 | 0.7234 | 0.9104 | 1.7234 | 1.7766 |
| | Shanghai | 0.0005876 | 0.7707 | 0.8570 | 1.7707 | 1.7293 |
| | Hangzhou | 0.0003653 | 0.8730 | 0.9291 | 1.8730 | 1.6270 |
| **South-west** | Chongqing | 0.0003146 | 0.8115 | 0.8455 | 1.8115 | 1.6885 |
| | Chengdu | 0.0003350 | 0.8697 | 0.8206 | 1.8697 | 1.6303 |

**Note**: Based on power law distribution, the self-similar fractal dimension $D_f$ can be converted into self-affine fractal dimension $D_s$, and *vice versa*.

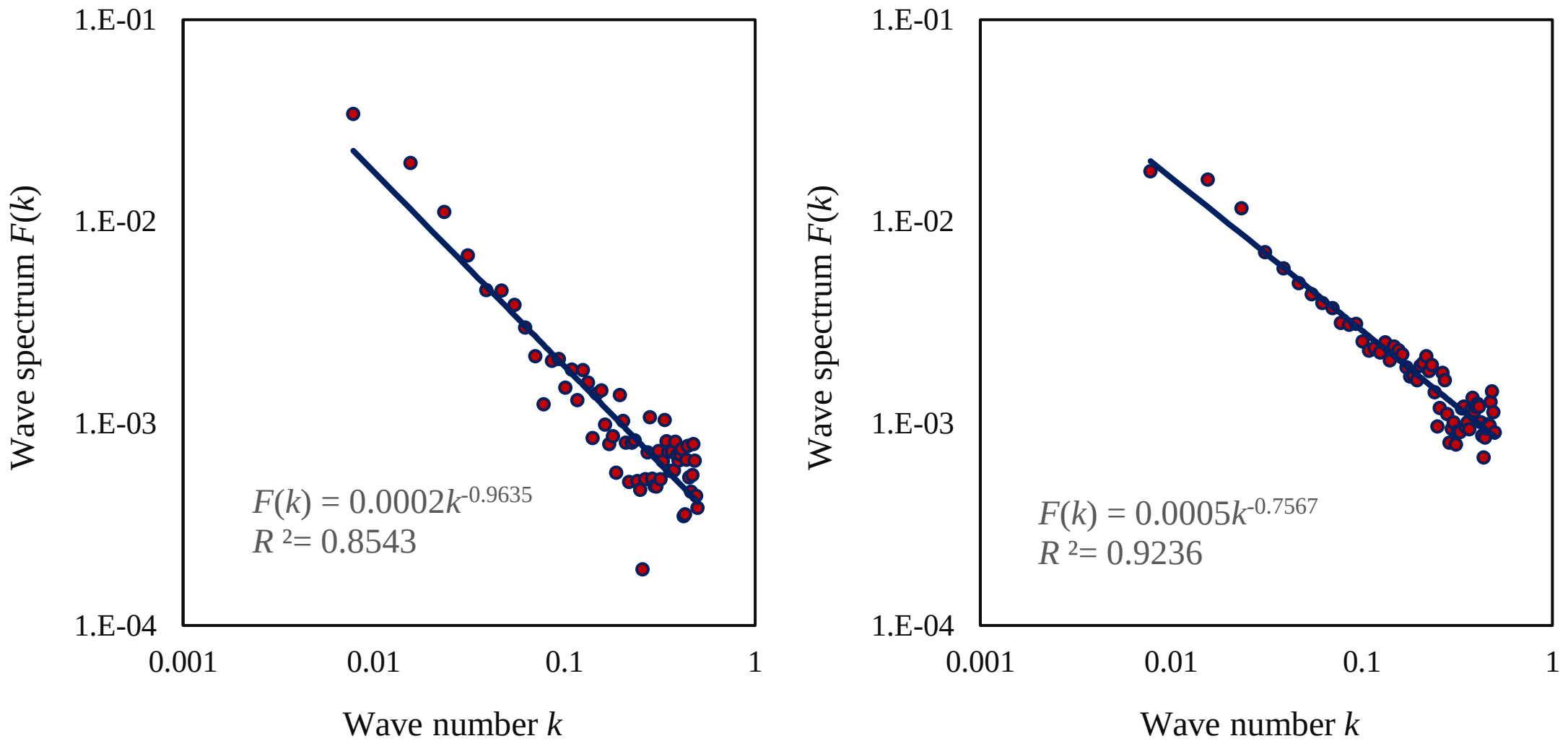

a. Beijing

b. Tianjin

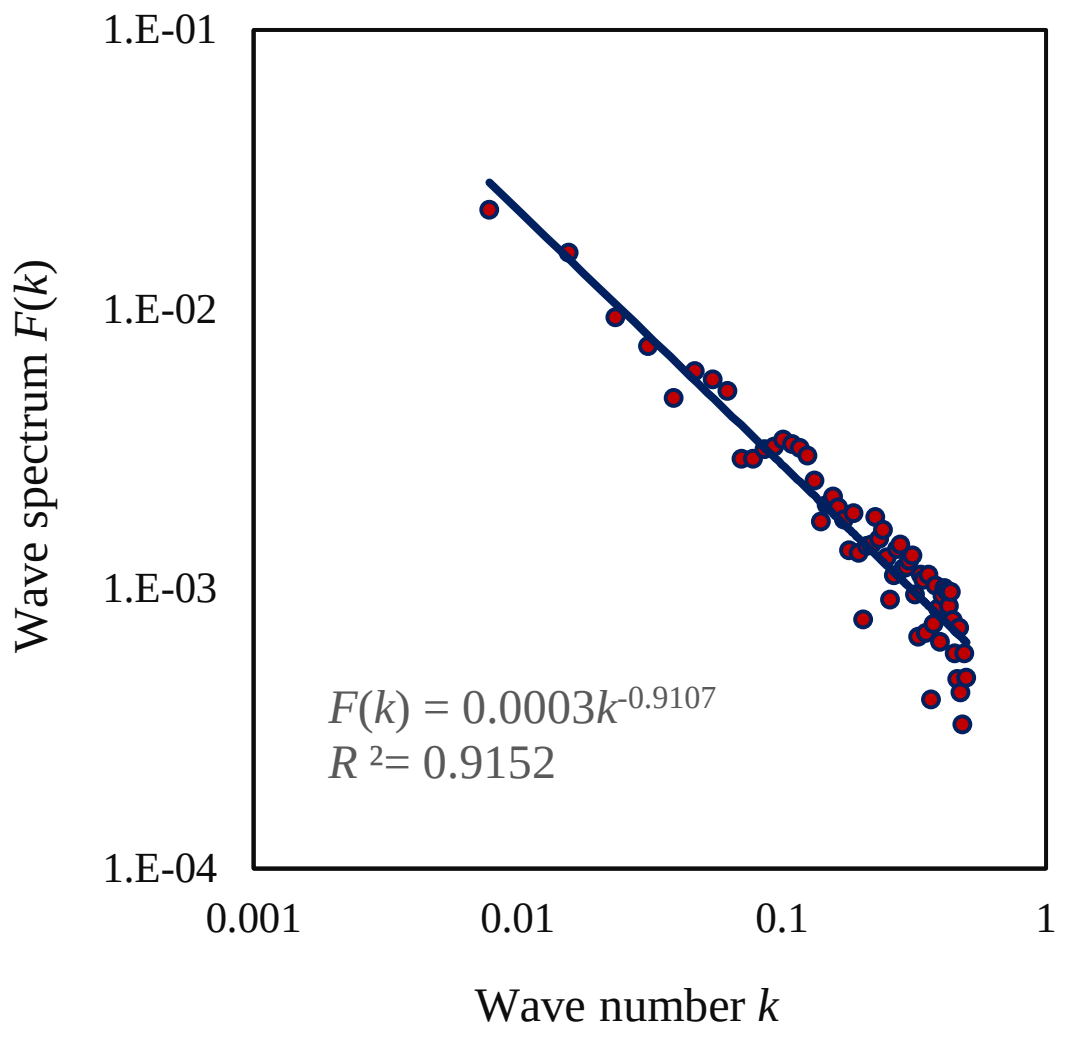

c. Guangzhou

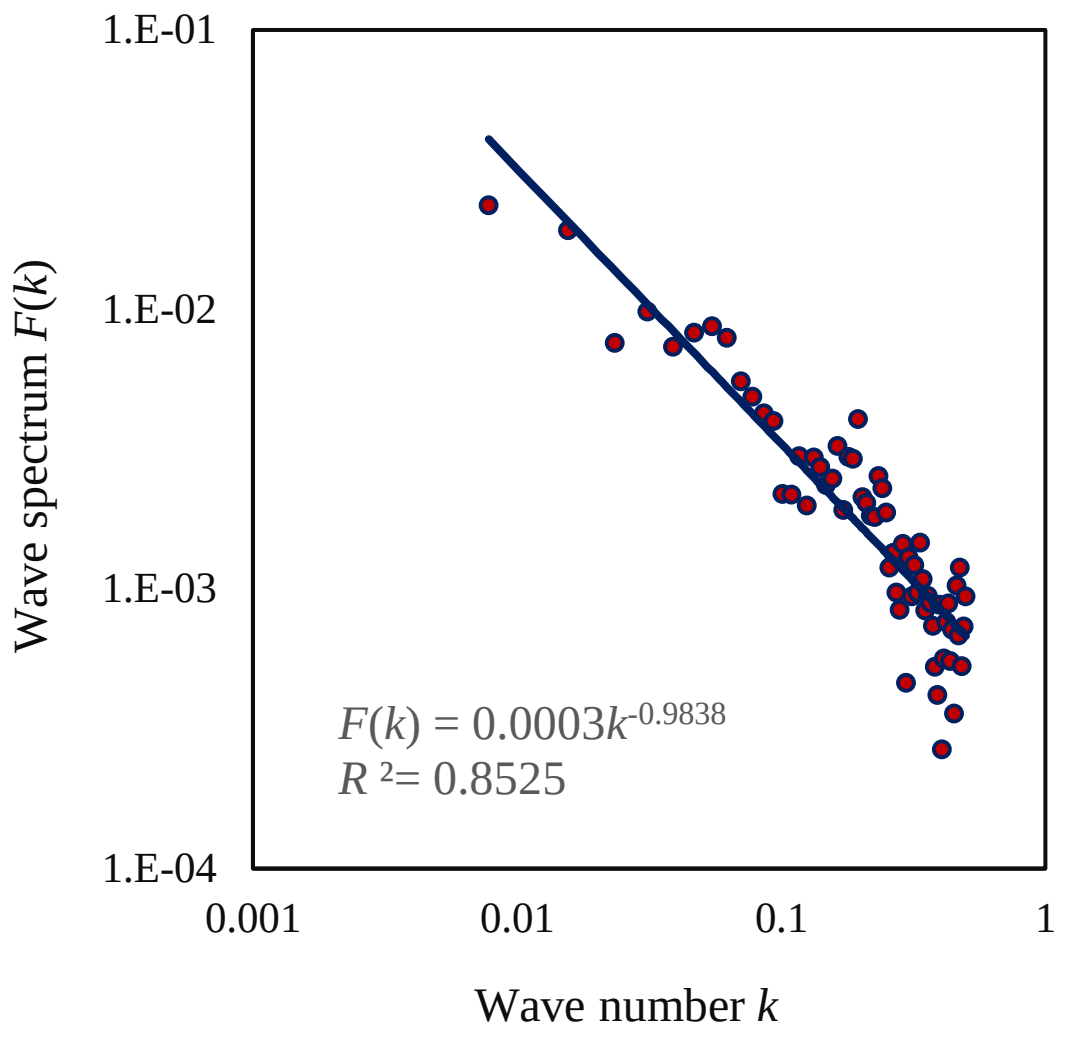

d. Shenzhen

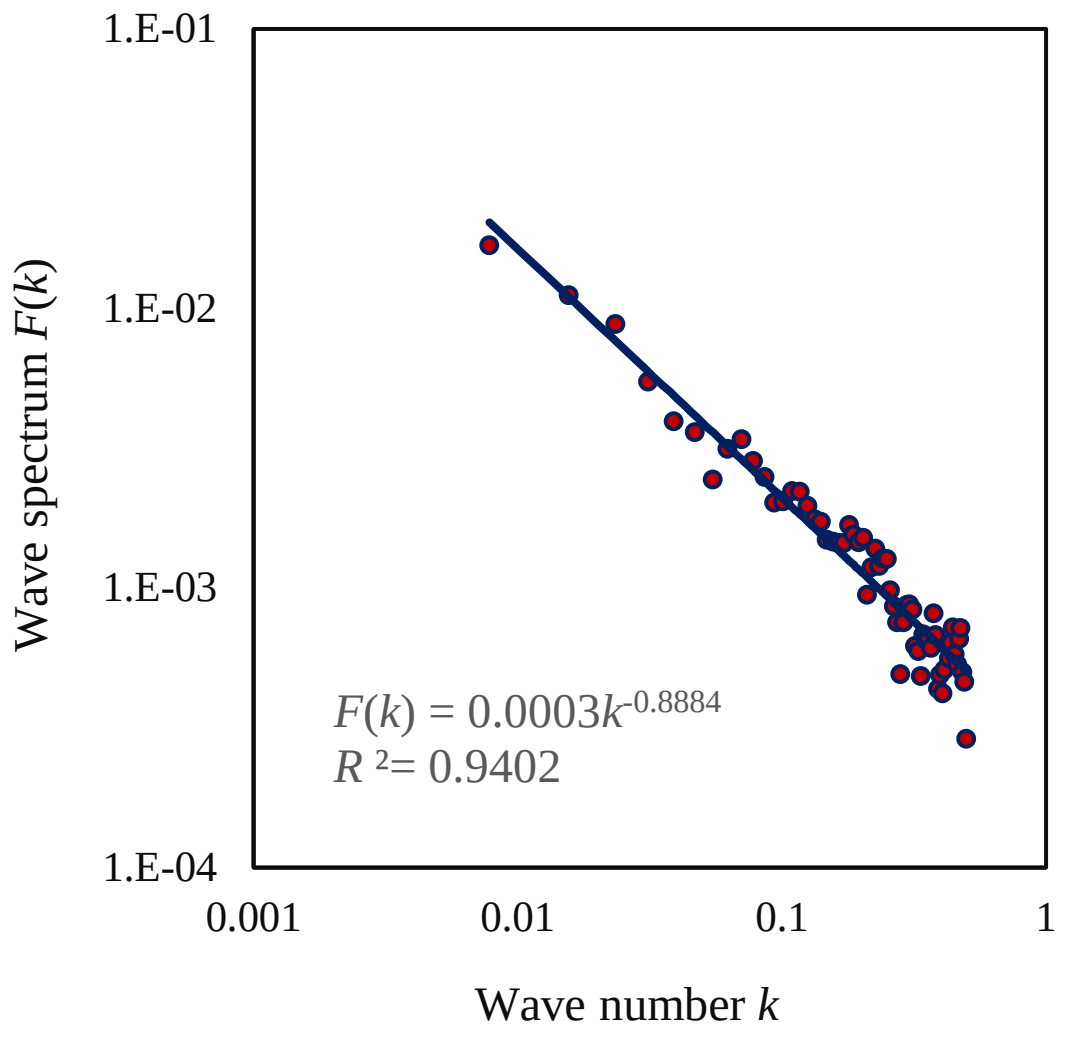

e. Wuhan

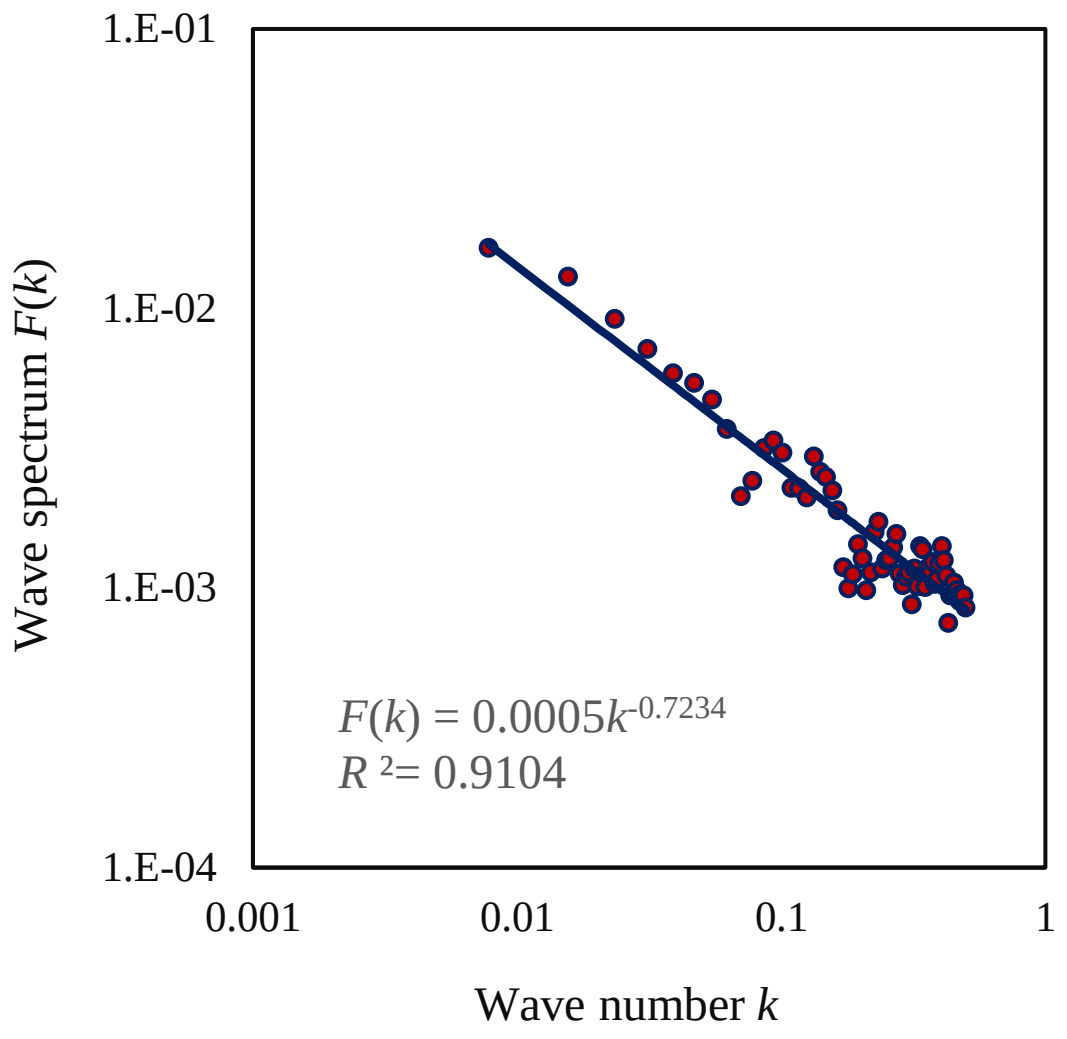

f. Nanjing

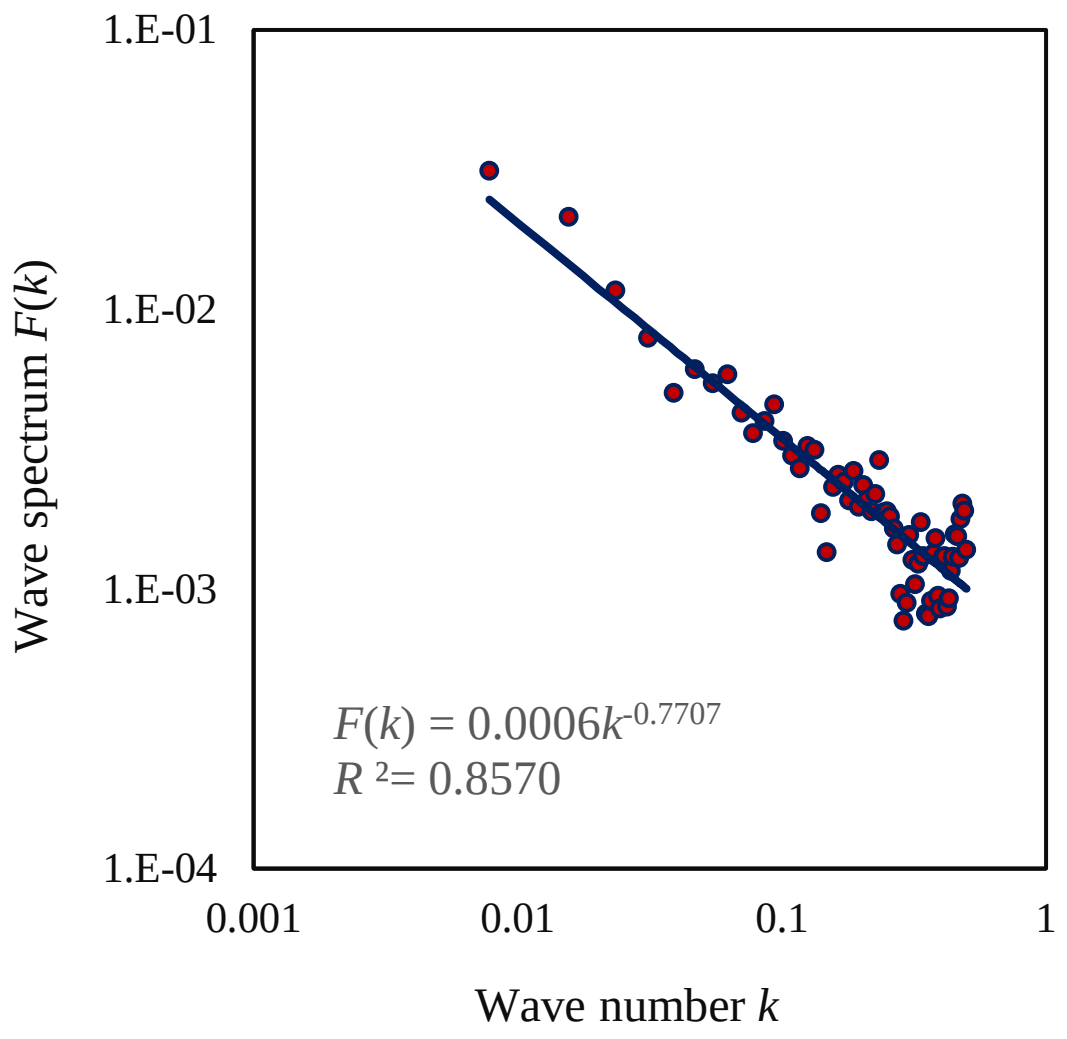

g. Shanghai

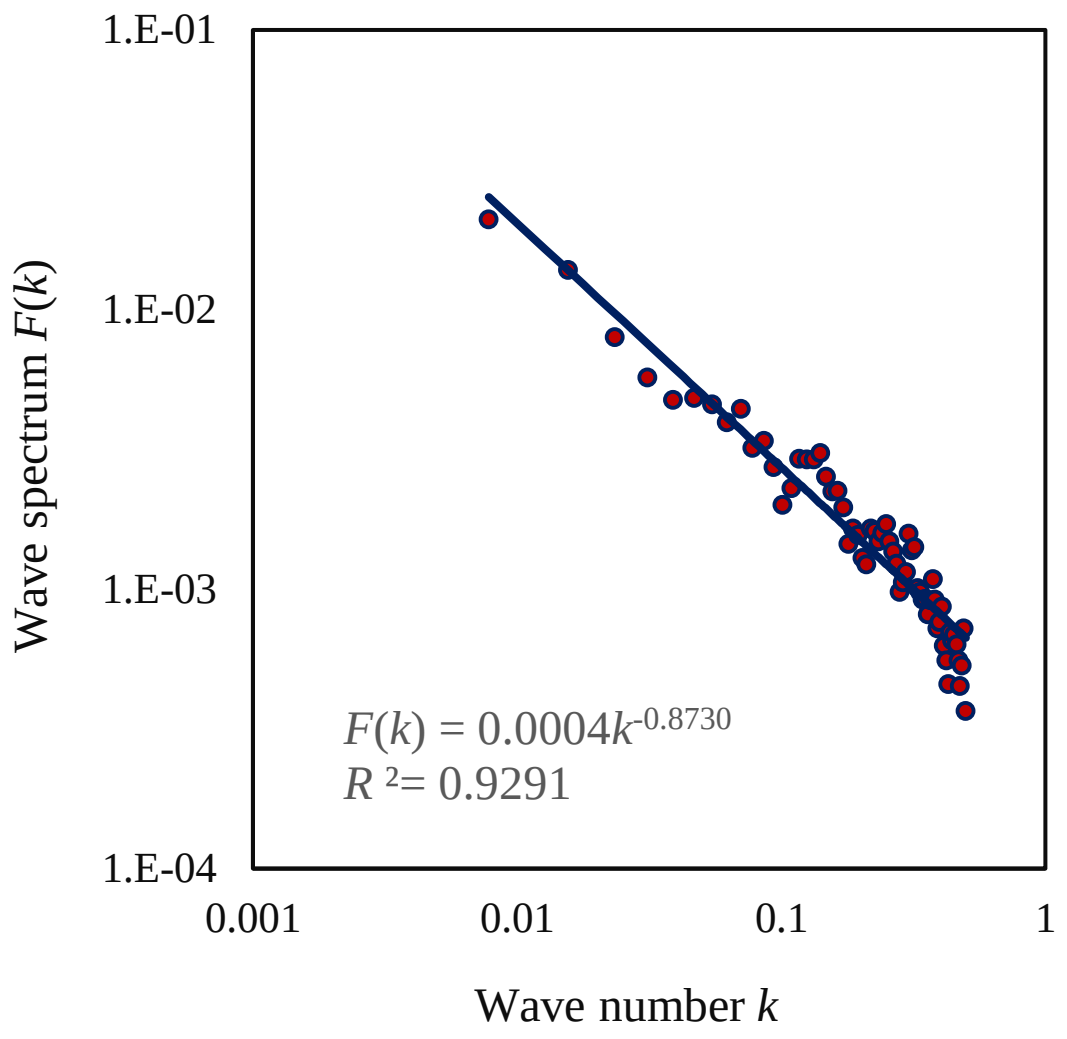

h. Hangzhou

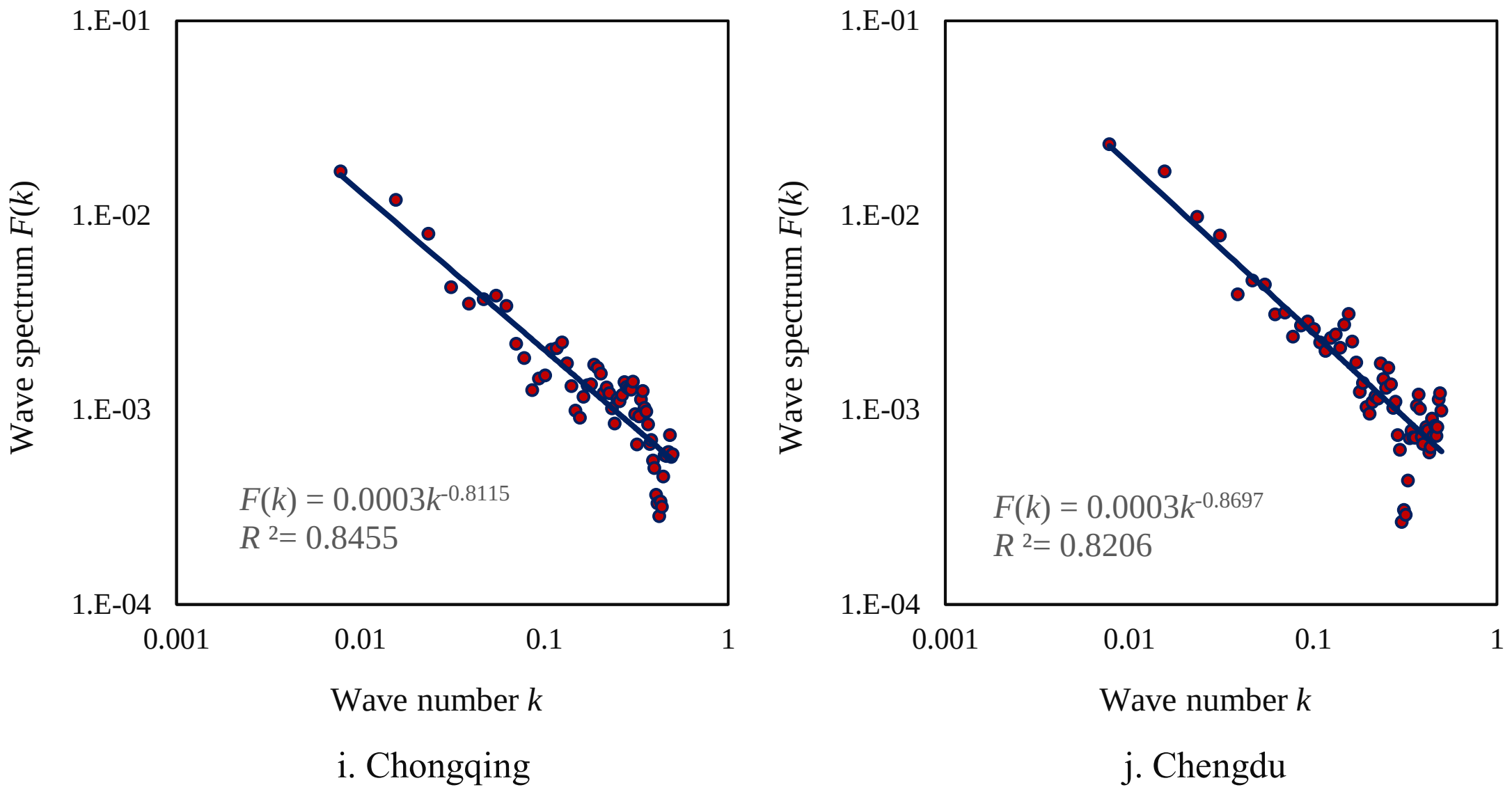

**Figure 5 The wave-spectrum scaling relations between wave numbers and wave spectrums (2016)**

**Note**: All the spectral exponents were estimated by equation (13), which is based on equation (6). By using equation (8), we can turn the spectral exponent values into fractal dimension values easily.

The fractal dimension estimated by wave-spectrum scaling can be used to replace radial dimension. The radial dimension has three meanings for spatial analyses. First, degree of space filling. The denser the traffic network of a city, the higher the fractal dimension value. Second, degree of spatial homogeneity. The slower the attenuation of traffic network density from a city center to the periphery, the higher the fractal dimension value. Third, degree of spatial correlation. The stronger the accessibility from one place to another, the higher the fractal dimension value. Beijing's road network has higher degrees of space filling, spatial uniformity, and spatial correlation, its radial dimension is high. Tianjin's network density is lower than Beijing's network density, indicating lower spatial filling extent, and its fractal dimension is significantly lower than Beijing's. Shanghai's road network has lower spatial homogeneity, that is, the spatial decay rate of urban network density is higher. Therefore, the radial dimension of Shanghai's road network is not so high. The distribution of road network in Nanjing is similar to some extent to that of Shanghai. In Nanjing, higher network density attenuation rate leads to relatively low fractal dimension. The traffic network filling degree of Hangzhou is not as high as that of Shanghai, but the network attenuation rate is lower than that of Shanghai. Wuhan is the hub of land and water transportation in Central China. Probably due to the Yangtze River, the road network filling degree of Wuhan is not particularly high. The fractal dimension of Guangzhou road network is relatively high. The distribution of traffic network in Guangzhou is similar to a degree to that of Beijing. Shenzhen's road network bear lower space filling extent, but has higher spatial homogeneity. In other words, the spatial decay of Shenzhen's network density is not significant. Therefore, the radial dimension of Shenzhen's road network is very high. In southwest China, Chengdu is located in the plain area, so its road network bears higher fractal dimension. In contrast, Chongqing is located in the mountainous area, and its road network possesses relatively low fractal dimension. Mountains and rivers reduce the fractal dimension values of urban traffic networks because they affect the space-filling degree of streets and roads in urban regions.

# 4. Discussion

The empirical analyses show that the method of wave-spectrum scaling analysis based on Fourier transform can be employed to estimate fractal dimension of spatial signals. Using these fractal parameters, we can make spatial analyses by means of geographical signals. Spatial signal extraction is not limited to traffic networks, but involve many aspects of geographic systems. Note that fractal approach can only be applied to fractal phenomena. In fact, spatial signals can be divided into two categories and four subcategories (Table 3). If we meet stationary spatial series of geographical signals, the wave-spectrum scaling analysis will be useless. If we meet the trend series of spatial signals with characteristic lengths, generally speaking, the wave-spectrum scaling is useless. Sometimes, the wave-spectrum scaling analysis can be applied to the spatial series satisfying exponential distributions, but the spectral exponent indicates a Euclidean dimension (Chen, 2008; Liu and Liu, 1993). If a self-similar spatial signal series, i.e., the spatial series without characteristic length, is encountered, the method presented in this paper is suitable. If a self-affine spatial signal series appears, the spectral analysis can also be applied to it in a proper way. In particular, if the spatial patterns for signal extraction are not incomplete, the wave-spectrum scaling is indispensable for estimating radial dimension of growing fractal processes. Incomplete spatial patterns result in power law degeneration, and fractal dimension cannot be directly estimated by Smeed's model.

**Table 3 Different types of spatial signals and corresponding analytical methods**

| Category | Type | Example | Method |
|---|---|---|---|
| **Non-fractal** | Stationary series | Random spatial | Conventional spatial statistics |
| **signals** | | processes | |
| | Trend series with characteristic scale | Exponential distribution of urban population density | Conventional mathematical methods and wave-spectrum scaling |
| **Fractal signals** | Self-similar series | Isotropic growing fractal networks | Wave-spectrum scaling based on correlation function |
| | Self-affine series | Anisotropic growing fractal networks | Wave-spectrum scaling relation |

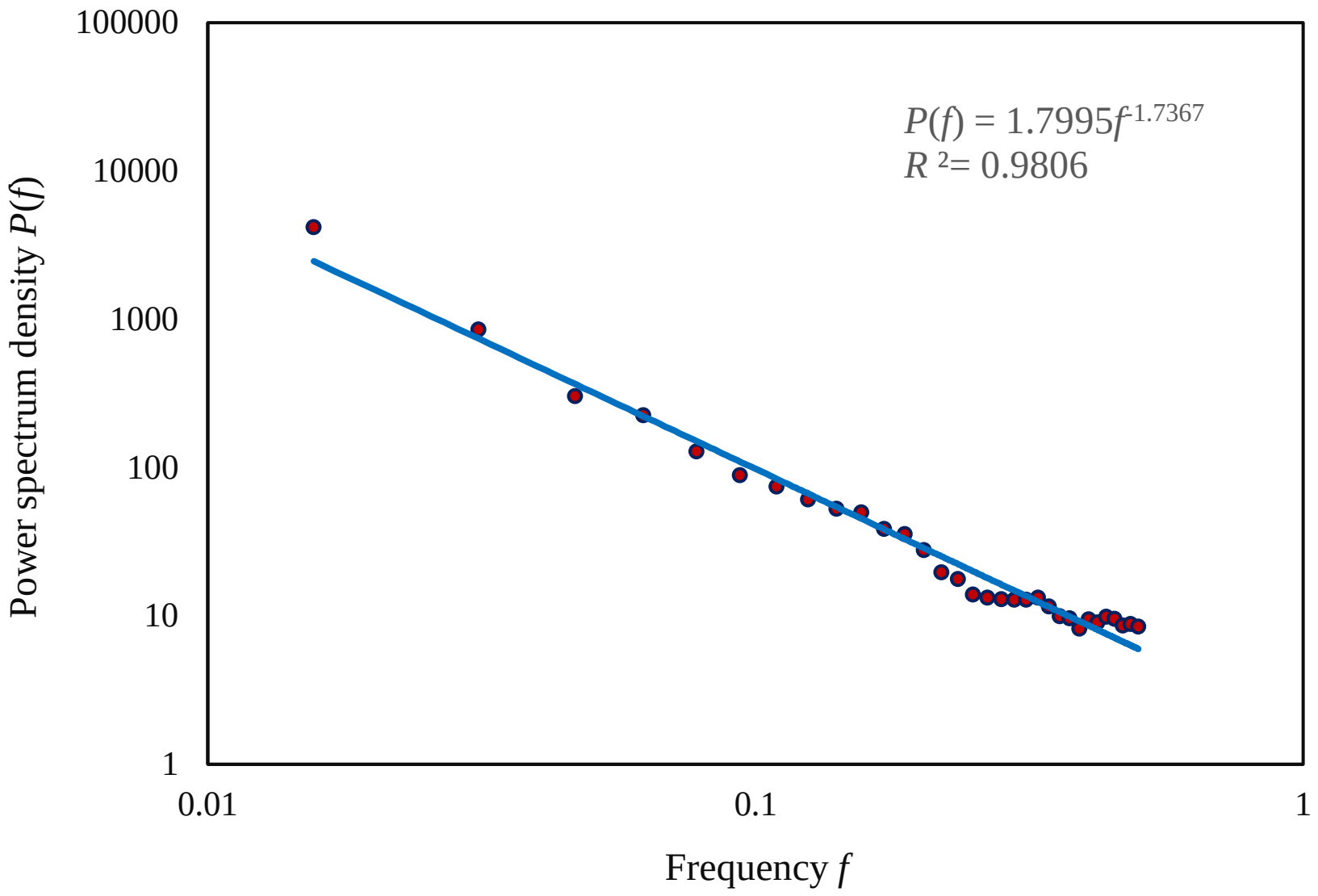

**Figure 6 The power-spectrum scaling relations between frequency and power spectral density (1956-2019)**

**Note**: The time series comes between 1949 and 2019. The number of data points is $N$=71. According to the requirement of FFT, the data point number must be $N=2^{j+1}$, where $j$=0, 1, 3 ,…int(ln$N$/ln(2))±1. The best number is $N$=64. So, the first 7 data points were removed to meet the needs of the algorithm.

In practice, the Fourier scaling relation can be equivalently replaced by the wave-spectrum scaling relation. The wave-spectral analysis can be applied to self-affine fractal structure of temporal and spatial signals (Chen, 2008). Traditionally, the wave-spectral function is defined as follows

$$W(k)=\frac{|F(k)|^2}{N}=\frac{F(k)\overline{F(k)}}{N}=\frac{a^2+b^2}{N}, \tag{19}$$

where $W(k)$ denotes wave-spectrum density. For the self-affine fractal series, there is a scaling relation between wave-spectrum density and wave number, that is

$$W(k)=W_0 k^{-\beta}, \tag{20}$$

where $\beta$ is spectral exponent, and $W_0$ refers to the proportionality coefficient. It can be proved that $\beta$=2$\alpha$. Thus the self-affine record dimension can be expressed as (Chen, 2010; Feder, 1988; Liu and Liu, 1993; Takayasu, 1990)

$$D_s = \frac{5-\beta}{2} = \frac{5}{2} - \alpha, \tag{21}$$

Accordingly, the self-similar trail dimension can be calculated by (Chen, 2010; Chen, 2013)

$$D_f = \frac{7}{2} - D_s = \frac{\beta}{2} + 1 = \alpha + 1. \tag{22}$$

That is to say, a self-affine fractal can be approximately described with self-similar fractal dimension. Self-similar process and self-affine process are related to and different from one another. In fractal theory, self-similarity is a special case of self-affinity. The former means isotropic growth, and the latter means anisotropic growth. However, for given direction, a self-affine process can be treated as self-similar process.

Complex signal analysis based on time series give self-affine record fractal dimension. Despite the similarity in mathematical essence, the form differs from spatial signal analysis. Replacing wave number $k$ and wave spectral density $W(k)$ with frequency $f$ and power spectral density $P(f)$, we can turn equation (20) into a frequency-spectrum relation as below

$$P(f) = P_0 f^{-\beta}, \tag{23}$$

in which $P_0$ denotes the proportionality coefficient. Equation (23) can be used to analyze temporal signals through fractal parameters. For example, applying equation (23) to the time series of urbanization level of China yields a model as follows

$$\hat{P}(f) = 1.7995 f^{-1.7367}. \tag{24}$$

The goodness of fit is about $R^2$=0.9806, the power spectral exponent is about $\beta$=1.7367 (Figure 6). According to equation (21), the fractal dimension of self-affine record is about $D_s$= (5-1.7367)/2 =1.6316. Thus the Hurst exponent is $H$=2-$D_s$=0.3684. This suggests an anti-persistence process of urbanization. Urbanization contains fluctuation process or negative feedback process. In fact, urbanization process contains urban systems and urban form (Knox and Marston, 2009). The frequency-spectral scaling can be applied to urban form and growth in an urban systems by both temporal and spatial signals. For instance, based on the sample paths of nighttime light (NTL) time series (1992-2013), we can calculate the spectral exponent of urban growth of the cities in Beijing-Tianjin-Hebei region of China (Figure 7, Table 4). The NTL data were originally processed for multi-scaling allometric analyses (Long and Chen, 2019b). The spectral exponents can be converted into self-affine record dimension and self-similar trail dimension, and associated with Hurst exponent based on R/S analysis (Hurst *et al*, 1965; Mandelbrot, 1982). The fractal dimension belong to phase space. The results suggest that two types of cities possess higher self-similar fractal dimension values. One is the large cities such as Beijing and Tianjin, and the other is the cities of self-affine growth such as Zhangjiakou and Qinhuangdao. The Hurst exponent values show that all the growing processes of all these cities bear long memory, moreover, all growing processes of all these cities but Handan bear persistence.

**Table 4 The power spectrum exponents, fractal dimension, and Hurst exponent of Chinese cities in Beijing-Tianjin-Hebei region based on time series of night light data (1992-2013)**

| City | Spectral analysis | | | | | R/S analysis | | |
|---|---|---|---|---|---|---|---|---|
| | $P_0$ | $\beta$ | $R^2$ | $D_s$ | $D_f$ | Coefficient | $H$ | $R^2$ |
| **Baoding** | 1532.2018 | 1.3649 | 0.5605 | 1.8176 | 1.6824 | 1.0111 | 0.5982 | 0.9373 |
| **Beijing** | 25950.7422 | 1.8009 | 0.8385 | 1.5995 | 1.9005 | 0.9317 | 0.6642 | 0.9542 |
| **Cangzhou** | 1545.2501 | 1.3164 | 0.5322 | 1.8418 | 1.6582 | 0.9809 | 0.6280 | 0.9696 |
| **Chengde** | 194.4615 | 1.4499 | 0.7281 | 1.7750 | 1.7250 | 0.9598 | 0.6366 | 0.9152 |
| **Handan** | 1680.5985 | 1.4165 | 0.4976 | 1.7917 | 1.7083 | 1.2221 | 0.4628 | 0.7739 |
| **Hengshui** | 295.9529 | 1.2741 | 0.6029 | 1.8629 | 1.6371 | 0.9621 | 0.6091 | 0.9344 |
| **Langfang** | 2493.1480 | 1.4288 | 0.6164 | 1.7856 | 1.7144 | 1.0578 | 0.5842 | 0.9222 |
| **Qinhuangdao** | 447.7705 | 1.6063 | 0.7843 | 1.6968 | 1.8032 | 1.1647 | 0.5625 | 0.9301 |
| **Shijiazhuang** | 1866.2198 | 1.5853 | 0.6874 | 1.7074 | 1.7926 | 0.9727 | 0.6074 | 0.9366 |
| **Tangshan** | 3534.1318 | 1.5370 | 0.7094 | 1.7315 | 1.7685 | 1.0489 | 0.5620 | 0.9332 |
| **Tianjin** | 24914.8032 | 1.5869 | 0.7798 | 1.7065 | 1.7935 | 1.0585 | 0.5543 | 0.9487 |
| **Xingtai** | 434.9114 | 1.5473 | 0.6817 | 1.7264 | 1.7736 | 0.9637 | 0.5869 | 0.9412 |
| **Zhangjiakou** | 410.9478 | 1.6865 | 0.7945 | 1.6568 | 1.8432 | 1.0666 | 0.5908 | 0.9131 |

Fourier transform and frequency/wave-spectrum scaling is one of important approaches for spatial analysis of geographical fractals. In urban studies, this method is useful for indirectly estimation of fractal dimension, especially when fractal structure is hidden by random noises. Compared with previous research, this work bears clear novelty. First, the previous works are devoted to theoretical models and fractal parameter relations (Chen, 2010). In particular, using wave-spectral scaling analysis, new fractal parameter relations are derived (Chen, 2013). In contrast, this work is devoted to empirical analysis of spatial signals. Second, in partial previous works are devoted to self-affine fractal analysis (Chen, 2008). Especially, the relationships between self-affine fractals and self-similar fractals are revealed by means of wave-spectral scaling relations. In contrast, this paper is chiefly devoted to self-similar fractal series. Third, in previous studies, the cases are based on land use patterns of cities (Chen, 2013; Chen, 2019). In contrast, this paper is devoted to research urban traffic networks. The shortcomings of this studies lies in two respects. First, this paper is actually devoted to 1-dimensional spatial signals. In other words, with the help of statistical average, the 2-dimensional spatial signals are converted into 1-dimensional spatial signals. The 2-dimensional random signals in strict sense have not been processed. Second, this paper is principally devoted to self-similar fractal signals. For the self-affine fractal signals, no deep analysis is made. There are too many problems to be solved for self-affine fractal signals. Limited to space, the relevant issues will be discussed in the future.

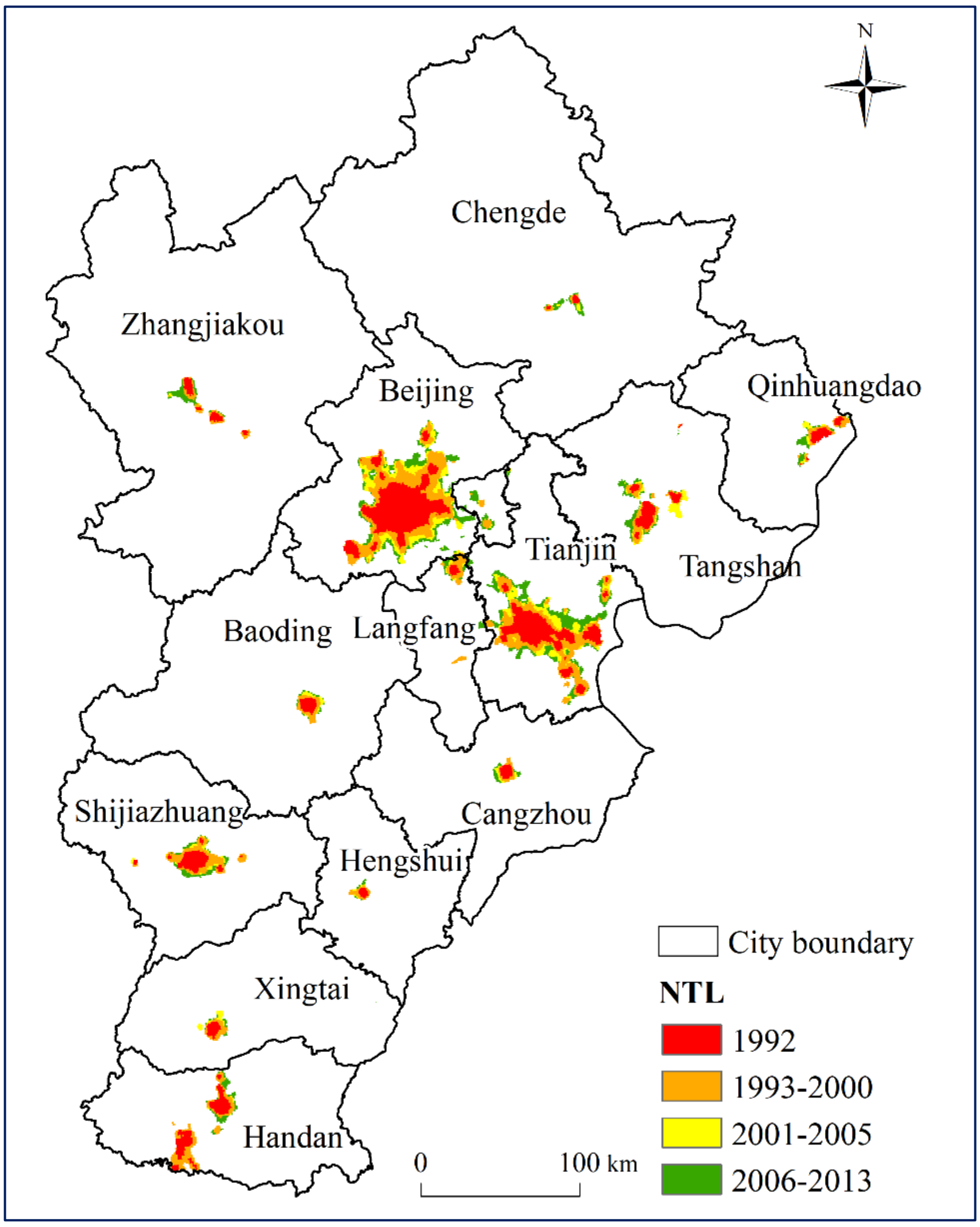

**Figure 7 Urban growth in Beijing-Tianjin-Hebei region reflected by night light data (1992-2013)**

**Note**: The changing process of total quantity of nighttime lights (NTLs) reflects urban growth. Using time series of night light data, we can make power spectral analysis and estimate the self-affine record dimension for temporal signals. A set of temporal signals reflect spatial signals from cross-sectional angle of view.

# 5. Conclusions

Fourier transform and wave-spectrum scaling relation can be applied to fractal-based spatial signal analysis. The main points of this paper are as follows. **First, if the self-similar fractal structure of spatial signals is concealed by random noises, or spatial patterns are not incomplete for signal extraction, the wave-spectrum scaling relation can be utilized to estimate fractal dimension**. Turning spatial signals into density distribution, we can make Fourier transform for the density series. The results can be converted into wave spectrum or spectral density series. Then, using the power law relation between wave number and wave spectrum, we can calculate spectral scaling exponent. The spectral exponent can be converted to fractal dimension, which can be employed to make spatial analysis for geographical signals. **Second, if spatial signals bear self-affine fractal structure, we can use wave-spectrum scaling to compute self-affine record dimension**. Then, the anisotropic fractal process can be treated as isotropic fractal process approximately. Thus, the fractal dimension of self-affine records can be converted to the fractal dimension of self-similar trails. Combining the self-affine fractal parameter and self-similar fractal parameter, we can make spatial analysis for geographical signals. **Third, the wave-spectrum scaling can be naturally applied to temporal signals of geographical systems**. If we have a time series representing temporal signals of geographical systems, e.g., the level of urbanization, we can apply the wave-spectrum scaling to the dataset. If the temporal signal bears fractal property, we will be able to work out the self-affine record dimension. By means of the fractal dimension of self-affine record, we can make time series analysis for the temporal signal of urban evolution.

## Acknowledgements

This research was sponsored by the National Natural Science Foundations of China (Grant No. 41671167). The support is gratefully acknowledged. The nighttime light dataset (1992-2014) come from American NOAA National Centers for Environmental Information (NCEI).

# References

Batty M, Longley PA (1994). *Fractal Cities: A Geometry of Form and Function*. London: Academic

Press
Brockwell PJ, Davis RA (1998). *Time Series: Theory and Methods (2ed)*. Berlin: Springer
Casti JL (1996). *Would-Be Worlds: How Simulation Is Changing the Frontiers of Science*. New York: John Wiley and Sons
Chen YG (2008). A wave-spectrum analysis of urban population density: entropy, fractal, and spatial localization. *Discrete Dynamics in Nature and Society*, vol. 2008, Article ID 728420, 22 pages
Chen YG (2010). Exploring the fractal parameters of urban growth and form with wave-spectrum analysis. *Discrete Dynamics in Nature and Society*, Volume 2010, Article ID 974917
Chen YG (2011). *Mathematical Methods for Geography*. Beijing: Science Press [In Chinese]
Chen YG (2013). Fractal analytical approach of urban form based on spatial correlation function. *Chaos, Solitons & Fractals*, 49(1): 47-60
Chen YG (2014). The spatial meaning of Pareto's scaling exponent of city-size distributions. *Fractals*, 22(1-2):1450001
Chen YG (2019). Fractal dimension analysis of urban morphology based on spatial correlation functions. In: D'Acci L(ed.). *Mathematics of Urban Morphology*. Birkhäuser: Springer Nature Switzerland AG, pp21-53
Chen YG (2020). Equivalence relation between normalized spatial entropy and fractal dimension. *Physica A*, 553: 124627
Chen YG, Wang JJ, Feng J (2017). Understanding the fractal dimensions of urban forms through spatial entropy. *Entropy*, 19, 600
Clark C (1951). Urban population densities. *Journal of Royal Statistical Society*, 114(4): 490-496
Cramer F (1993). *Chaos and order: the complex structure of living systems* (translated by D.I. Loewus). New York: New York: VCH Publishers
Feder J (1988). *Fractals*. New York: Plenum Press
Frankhauser P (1998). The fractal approach: A new tool for the spatial analysis of urban agglomerations. *Population: An English Selection*, 10(1): 205-240
Frankhauser P, Sadler R (1991). Fractal analysis of agglomerations. In: Hilliges M (ed.). *Natural Structures: Principles, Strategies, and Models in Architecture and Nature*. Stuttgart: University of Stuttgart, pp 57-65
Harvey D (1969). *Explanation in Geography*. London: Edward Arnold Ltd.

Hurst HE, Black RP, Simaika YM (1965). *Long-term Storage: An Experimental Study*. London: Constable

Knox PL, Marston SA (2009). *Places and Regions in Global Context: Human Geography (5th Edition)*. Upper Saddle River, NJ: Prentice Hall

Liu SD, Liu SK (1993). *An Introduction to Fractals and Fractal Dimension*. Beijing: China Meteorological Press [In Chinese]

Long YQ, Chen YG (2019a). Fractal characterization of structural evolution of Beijing, Tianjin and Hebei transportation network. *Human Geography*, 34(4): 115-125 [In Chinese]

Long YQ, Chen YG (2019b). Multi-scaling allometric analysis of the Beijing-Tianjin-Hebei urban system based on nighttime light data. *Progress in Geography*, 38(1): 88-100 [In Chinese]

Mandelbrot BB (1982). *The Fractal Geometry of Nature.* New York: W. H. Freeman and Company

Pincus SM (1991). Approximate entropy as a measure of system complexity. *PNAS*, 88(6): 2297–2301

Ryabko BYa (1986). Noise-free coding of combinatorial sources, Hausdorff dimension and Kolmogorov complexity. *Problemy Peredachi Informatsii*, 22: 16-26

Shannon CE (1948). A mathematical theory of communication. *Bell System Technical Journal*, 27(3): 379-423

Smeed RJ (1963). Road development in urban area. *Journal of the Institution of Highway Engineers*, 10(1): 5-30

Sparavigna AC (2016). Entropies and fractal dimensions. *Philica*, hal-01377975

Takayasu H (1990). *Fractals in the Physical Sciences*. Manchester: Manchester University Press

White R, Engelen G (1993). Cellular automata and fractal urban form: a cellular modeling approach to the evolution of urban land-use patterns. *Environment and Planning A*, 25(8): 1175-1199

Zmeskal O, Dzik P, Vesely M (2013). Entropy of fractal systems. *Computers & Mathematics with Applications*, 66 (2): 135-146